\documentclass[twoside,twocolumn,english,superscriptaddres,showpacs,longbibliography,aps,prl]{revtex4-2}

\usepackage[T1]{fontenc}
\usepackage[utf8]{inputenc}
\usepackage[english]{babel}
\usepackage{bm}
\usepackage{amsmath}
\usepackage{amssymb}
\usepackage{graphicx}
\usepackage{circuitikz}
\usepackage[unicode=true, pdfusetitle, bookmarks=true,bookmarksnumbered=true,bookmarksopen=true,bookmarksopenlevel=1,
breaklinks=false,pdfborder={0 0 0},pdfborderstyle={},backref=false,color links, citecolor=blue,linkcolor=red,urlcolor=blue] {hyperref}
\usepackage[nameinlink]{cleveref}
\Crefname{figure}{Fig.}{}

\makeatletter
\def\set@firstnote#1{%
 \@ifnum{\firstnote@num=#1\relax}{}{%
  \class@warn@end{Endnote numbers changed: rerun LaTeX}%
 }%
 \immediate\write\@mainaux{%
   \global\mathchardef\string\firstnote@num#1\relax
 }%
}%

\usepackage{braket}
\usepackage{bm}
\usepackage{tikz}
\usepackage{pgfplots}
\usepackage{pgf}
\usepackage{subfigure}
\usepackage{nicematrix}
\usepackage{diagbox}
\usepackage{orcidlink}

\renewcommand{\selectlanguage}[1]{}

\makeatother

\pgfplotsset{compat=1.18}
\begin{document}

\title{Generalizing fusion rules by shuffle: Symmetry-based classifications of nonlocal systems constructed from similarity transformations}
\author{Yoshiki Fukusumi}
\affiliation{Physics Division, National Center of Theoretical Sciences, National Taiwan University, Taipei 106319, Taiwan}
\author{Taishi Kawamoto}
\affiliation{Center\! for Gravitational\! Physics\! and Quantum \! Information, Yukawa\! Institute\! for\! Theoretical\! Physics, Kyoto\! University, Kitashirakawa\! Oiwakecho, Sakyo-ku, Kyoto 606-8502, Japan}
\date{\today}
\begin{abstract}

We study fusion rings, or symmetry topological field theories (SymTFTs), which lie outside the non-negative integer matrix representation (NIM-rep), by combining knowledge from generalized symmetry and that from pseudo-Hermitian systems. By applying the Galois shuffle operation to the SymTFTs, we reconstruct fusion rings that correspond to nonlocal CFTs constructed from the corresponding local nonunitary CFTs by applying the similarity transformations. The resultant SymTFTs are outside of NIM-rep, whereas they are ring isomorphic to the NIM-rep of the corresponding local nonunitary CFTs. We study the consequences of this correspondence between the nonlocal unitary model and local nonunitary models. We demonstrate the correspondence between their classifications of massive or massless renormalization group flows and the discrepancies between their boundary or domain wall phenomena. Our work reveals a new connection between ring isomorphism and similarity transformations, providing the fundamental implications of ring-theoretic ideas in the context of symmetry in physics.

\end{abstract}

\begin{flushright}
YITP-25-181
\\
\end{flushright}
\maketitle

\section{introduction}
\label{introduction}

Similarity transformation, $\eta$, is a fundamental transformation, which enables one to obtain the Hermitian systems from pseudo-Hermitian systems by the \emph{invertible} transformations \cite{Mostafazadeh:2001jk,Mostafazadeh:2001nr,Mostafazadeh:2002id}: \begin{equation}
    H' = \eta H \eta^{-1}.  
\end{equation}
where $H$ is the Hamiltonian of the original pseudo-Hermitian systems and $H'$ is that of the corresponding Hermitian system\footnote{In this manuscript, we assume that the spectrum of pseudo-Hermitian systems are real, but it will be possible to relax this assumption to the reality of partition functions.}. This transformation also implements a dual relationship between the nonunitary conformal field theories (CFTs) and the nonunitary CFTs describing the corresponding lattice models \cite{Guruswamy:1996rk,Hsieh:2022hgi}. One can see a similar (or the same) dual relationship in  \cite{Buican:2017rya,Ferrari:2023fez,Creutzig:2024ljv} related to the $3d-3d$ correspondence \cite{Chung:2014qpa,Chun:2019mal,Cheng:2022rqr}. It should be remarked that the similarity transformation itself is an intrinsically quantum operation, because it changes the role of vacuum in a CFT and the ground state (or effective vacuum) at the level of linear algebra\cite{Gannon:2001uy,Gannon:2003de}. Because of this unusual aspect, it is difficult to represent the transformation in the standard Lagrangian formalism of quantum field theories (QFTs). Related to the property of the effective vacuum or ground state, the Zamolodchikov $c$-theorem in unitary CFTs \cite{Zamolodchikov:1986gt} has been generalized to $c_{\text{eff}}$-theorem \cite{Castro-Alvaredo:2017udm} with a connection to the $PT$ symmetry in non-Hermitian systems \cite{Bender:1998ke,Bender:1998gh}. 

One can notice the nontrivial relationship between pseudo-Hermitian systems and the corresponding \emph{nonlocal} Hermitian systems in the older studies on the realizations of chiral fermions \cite{Wilson:1974sk,Drell:1976mj,Susskind:1976jm,PhysRevD.26.468,Stamatescu:1993ga,Guo:2021sjp,Guo:2024jqt}. The lattice realization of the chiral fermions is restricted by the Nielsen-Ninomiya theorem \cite{Nielsen:1981hk,Nielsen:1981xu}, and the two possible ways to realize chiral fermions are breaking locality or Hermiticity. There also exists a lattice analogue in $1+1$-dimensions \cite{Castro-Alvaredo:2009xex}. We note a recent review on chiral fermion \cite{Ma:2024zbl} and those on non-Hermitian systems \cite{Bender:2007nj,Ashida:2020dkc,Okuma:2022bnb} as references. These research directions also imply the difficulty in describing the quantum models in the Lagrangian formalism of QFTs. We also note the pioneering nonlocal model, Haldane-Shastry model \cite{Haldane:1987gg,SriramShastry:1987wdh} and several recent research directions on non-local systems, numerical studies  \cite{Beenakker2023TangentFD,Haegeman:2024qgf,Zakharov:2024xcg}, long-range interacting integrable models  \cite{BenMoussa:2024xdm,Klabbers:2024abs}, extension of Lieb-Schultz-Mattis theorem \cite{doi:10.1142/S0217984996000080,TigranHakobyan_2003,Ma:2024sig,Zhou:2024whw,Liu:2024nox,Yang:2024rwz}, and quantum information scrambling \cite{Sachdev:1992fk,KitaevTalks,Kitaev:2017awl,Swingle:2016var,Wanisch:2022gyr,Tezuka:2022mrr,Bhattacharya:2023xjx,Nozaki:2013wia,Kawamoto:2022etl}. These studies have been developed in distinct research fields, with different techniques and languages. Hence, one may require some general method applicable to these quantum mechanical models, such as \emph{symmetry analysis} based on \emph{linear algebra}.  

On the other hand, there is a significant progress in the symmetry analysis in physics, known as generalized symmetry \cite{Gaiotto:2014kfa}. Roughly speaking, the symmetry in physics had been studied as a hierarchical structure of group, but this has been generalized to ring or algebra by treating the symmetry operators as conserved charges in quantum Hamiltonian systems\cite{Petkova:2000ip}. In more recent technical terms, the symmetry has often been represented as a fusion category, but the identification between a fusion category and a (fusion) ring is more involved through topological Wick rotation (or modular $S$ transformation). Implicitly, this correspondence requires mapping between the quantum and classical systems. The underlying subtleties of this mapping have been clarified more recently in  \cite{Yao:2020dqx} in studying representations of the parafermionic lattice models \cite{Fateev:1985mm}. Surprisingly, the parafermionic models represented by the parafermionic variables cannot be interpreted as two dimensional classical models, regardless of their apparent realization in the one dimensional quantum lattice models (equivalent to $Z_{N}$ generalization of Kitaev chain \cite{Kitaev:2001kla}). The resultant CFT can be interpreted as $Z_{N}$ extended models, and they are intrinsically quantum models. Moreover, the fusion category in literature usually does not permit noninteger coefficients before the anyonic objects, and this is intrinsically different from conserved charges of a quantum system, which should form a linear algebra. 

This subtlty does not affect the analysis for a model when restricting our attention to a well-organized model with quantum/classical correspondence, such as a bosonic model described by a modular tensor category. However, it is not the case when studying their hierarchical structure and related extended models, which are intrinsically quantum. The relationship between the extended model and hierarchical structures has been proposed in  \cite{Gaiotto:2012np} (One can see more detailed discussions and references in  \cite{Fukusumi:2024ejk,Fukusumi:2025ljx}). From this point, the sandwich construction or symmetry-based representation of the correspondence between CFT and topological quantum field theory (TQFT) \cite{Witten:1988hf} should provide boundary states (expressed as linear vectors) in $D+1$ dimensional TQFTs and symmetry operators in the $D$ dimensional CFTs \cite{Fukusumi:2023psx,Fukusumi:2024cnl,Fukusumi:2025ljx}. The resultant set of the conserved charges and their algebra have been known as symmetry topological field theories (SymTFTs) \cite{Apruzzi:2021nmk}. Related hierarchical structures have been studied as gapped domain wall \cite{Lan:2014uaa,Hung:2015hfa,kong2015boundarybulkrelationtopologicalorders,Kong:2017hcw,Zhao:2023wtg} in TQFTs or massless renormalization group flow \cite{Zamolodchikov:1987ti,Zamolodchikov:1987jf,Zamolodchikov:1989hfa} in CFTs.

In this paper, we construct and study a particular type of conserved charges, SymTFTs, of the nonlocal Hermitian systems derived from pseudo-Hermitian systems by the similarity transformations. For readers interested in SymTFTs and generalized symmetries, we note several earlier works \cite{article,Bockenhauer:1999wt,Petkova:2000ip,Fuchs:2002cm,Frohlich:2004ef,Cobanera:2009as,Freed:2009qp,Cobanera:2011wn,Cobanera:2012dc,Gaiotto:2012np,Gaiotto:2014kfa,Tachikawa:2017gyf,Bhardwaj:2017xup}, recent reviews \cite{McGreevy:2022oyu,Cordova:2022ruw,Bhardwaj:2023kri,Schafer-Nameki:2023jdn} and two works \cite{Fukusumi:2025clr,Fukusumi:2025xrj} by the authors and a collaborator. The fundamental algebraic structure is a generalization of the Verlinde formula  \cite{Gannon:2001uy,mueger2002structuremodularcategories,etingof2015tensor}, Eq. \eqref{effective_Verlinde_formula}, and the corresponding SymTFTs and their modules are written as Eq. \eqref{shuffle_SymTFT} and Eq. \eqref{shuffle_BCFT}. Their implications to RGs are summarized in Fig. \ref{fig:massless_flow}. However, the defect or boundary phenomena are more subtle because of the nonlocality, as we demonstrate in Sec. \ref{Difficulty_BCFT}.

\section{Nonunitary CFTs and their symmetry}
\label{Nonunitary}

First, we concentrate our attention on the $A$-type diagonal CFTs with the following partition functions,
\begin{equation}
Z=\sum_{\alpha} \chi_{\alpha} (\tau) \overline{\chi_{\alpha}}(\overline{\tau})
\end{equation}
where $\chi$ and $\overline{\chi}$ are the chiral or antichiral characters labelled by the primary field $\alpha$ and $\tau$ and $\overline{\tau}$ are the corresponding modular parameters. We note  \cite{DiFrancesco:1997nk,Recknagel:2013uja,Northe:2024tnm} as general references, and mainly follow the recent work \cite{Fukusumi:2025xrj} by the authors.

In a pseudo-Hermitian system, one can introduce the left and right energy eigenstates. Hence, corresponding to this fact, one can represent the characters as,
\begin{equation}
\chi_{\alpha}(\tau)= \sum_{M}\widetilde{\langle \alpha,M |}e^{2\pi i \tau (L_{0}-c/12)}| \alpha, M \rangle
\end{equation} 
where $\{ |\alpha, M\rangle \}$ is the right eigenstates of the Virasoro generator $L_{0}(\in \{ L_{m}\}_{m\in\mathbb{Z}})$ producing the corresponding Verma module labelled by the index $M$ of descendant fields, and the left eigenstates $\{ \tilde{\langle\alpha |} \}$ is its linear dual satisfying the relation $\widetilde{\langle\alpha, M |}|\beta, M' \rangle =\delta_{\alpha,\beta} \delta_{M,M'} $. We also note the form of the Virasoro algebra 
\begin{equation}
[L_{m},L_{n}]=(m-n)L_{m+n}+\frac{c}{12}(m^{3}-m)\delta_{m+n,0}
\end{equation}
where $c$ is the central charge and $\delta_{a,b}$ is the Kronekker delta.
One can also implement the chiral projection operators $P_{\alpha,M}=|\alpha,M \rangle\widetilde{\langle\alpha, M |}$. In a similar way, one can also implement their antichiral analogue $\overline{P_{\alpha,\overline{M}}}$. Hence, we introduce the following idempotent,
\begin{equation}
P_{\alpha}=\sum_{M,\overline{M}}P_{\alpha,M}\overline{P_{\alpha,\overline{M}}}
\end{equation} 
satisyfying the relation $P_{\alpha}P_{\beta}=\delta_{\alpha,\beta} P_{\alpha}$. In this expression, the unitary CFTs can be interpreted as exceptional models where the linear dual coincides with the complex conjugation, i.e. $\left(|\alpha,M\rangle\right)^{\dagger}=\langle \alpha,M |$.

For the latter convenience, we introduce the modular $T$ and $S$ transformations,
\begin{equation}
T: \tau \rightarrow \tau+1, \  S: \tau \rightarrow -1/\tau
\end{equation}
and their action to the characters,
\begin{align}
&T: \chi_{\alpha}(\tau +1)=e^{2\pi i \left(h_{\alpha}-\frac{c}{24}\right)}\chi_{\alpha}(\tau ) \\
&S: \chi_{\alpha}\left(-1/\tau\right)=\sum_{\beta} S_{\alpha,\beta} \chi_{\beta} (\tau ).
\end{align}
where $h$ represents the conformal dimension.

The Verlinde formula \cite{Verlinde:1988sn} is written as follows:
\begin{equation}
N^{\gamma}_{\alpha,\beta}=\sum_{\delta}\frac{S_{\alpha,\delta}S_{\beta,\delta}\overline{S_{\delta,\gamma}}}{S_{I,\delta}}
\label{Verlinde_formula}
\end{equation}
where $I$ is the identity operator corresponding to the vacuum and $N$ is the so-called nonnegative integer matrix (NIM), specifying the fusion rule $\alpha \times \beta =\sum_{\gamma} N^{\gamma}_{\alpha,\beta}\gamma$. We stress that even in nonunitary models, the matrix $N$ is NIM. In this setting, one can introduce the Verlinde line operator as follows \cite{Petkova:2000ip},
\begin{equation}
\mathcal{Q}_{\alpha}=\sum_{\beta} \frac{S_{\alpha,\beta}}{S_{I, \beta}}P_{\beta}
\end{equation}
where these operators also satisfy the relation $\mathcal{Q}_{\alpha}\times\mathcal{Q}_{\beta}=\sum_{\gamma}N^{\gamma}_{\alpha,\beta}\mathcal{Q}_{\gamma}$ by intepreting the symbol $\times$ as multiplication of operators. Remarkablly, $\mathbf{A}=\{ \mathcal{Q}_{\alpha}\}_{\alpha}$ is a set of linear operators over the complex number field $\mathbb{C}$ following the pioneering work \cite{Petkova:2000ip}. By definition, $\{ \mathcal{Q}_{\alpha}\}$ defined by the projections commutes with the CFT Hamiltonian, and one can interprete them as conseved charges or SymTFTs. For latter use, we also note a relation,
\begin{equation}
P_{\alpha}=\sum_{\beta}\mathcal{Q}_{\beta} S_{I,\alpha}\overline{S_{\alpha,\beta}}.
\end{equation}
Hence, $\{ P_{\alpha}\}$ and $\{ \mathcal{Q}_{\alpha}\}$ are ring isomorphic.

The form of $\mathcal{Q}_{\alpha}$  implies the existence of the modules (or smeared BCFTs \cite{Cardy:2017ufe,Thorngren:2019iar,Kikuchi:2022gfi,Kikuchi:2023gpj,Fukusumi:2024ejk,Fukusumi:2025clr,Fukusumi:2025xrj}) spanned by the left and right Cardy's states,
\begin{equation}
|\alpha\rangle =\sum_{\beta}\frac{S_{\alpha,\beta}}{\sqrt{S_{I,\beta}}}|\beta \rangle \rangle, \widetilde{\langle\alpha|} =\sum_{\beta}\frac{\overline{S}_{\alpha,\beta}}{\sqrt{S_{I,\beta}}}\widetilde{\langle \langle \beta |}
\end{equation}
where $\{ |\alpha\rangle\rangle\}$ is the Ishibashi states \cite{Ishibashi:1988kg} satisfying the relation, $\widetilde{\langle\langle \alpha|} e^{\pi i\tau (L_{0}+\overline{L_{0}}-c/12)} |\beta \rangle \rangle=\chi_{\alpha}(\tau)\delta_{\alpha,\beta}.$ We remind of the readers that the states and their linear duals are fundamental, and this is true for constructing the boundary states. By applying the Verlinde formula, the symmetry operators act on the module in the following way \cite{Graham:2003nc},
\begin{equation}
\mathcal{Q}_{\alpha}|\beta\rangle =\sum N_{\alpha,\beta}^{\gamma}|\gamma \rangle
\end{equation}
In other words, the boundary states are defined as a natural \emph{module} in linear algebra where the symmetry operators naturally act on. This aspect has been noticed in  \cite{Graham:2003nc}, and related phenomenologies have been developped in  \cite{Cardy:2017ufe,Thorngren:2019iar,Kikuchi:2022gfi,Kikuchi:2023gpj,Fukusumi:2024ejk,Fukusumi:2025clr,Fukusumi:2025xrj}.
Moreover, the Cardy states satisfy the following relation \cite{Cardy:1986gw,Cardy:1989ir},
\begin{equation}
Z_{\alpha,\beta}(\tau)=\widetilde{\langle\alpha|}e^{2\pi i (L_{0}+\overline{L_{0}}-c/12)}|\beta \rangle=\sum_{\gamma}N^{\gamma}_{\alpha,\beta}\chi_{\gamma}(-1/\tau)
\end{equation}

As can be seen in Eq.\eqref{Verlinde_formula}, the vacuum, $I$, plays the central role. However, in nonunitary CFTs, there exists another fundamental object, the effective vacuum, $o$ where the conformal dimension $h_{o}$ is lowest in the theory. Hence, as an analog of local unitary RCFTs, one can naively expect the following to have some meaning,
\begin{equation}
N^{\gamma, [o]}_{\alpha,\beta}=\sum_{\delta}\frac{S_{\alpha,\delta}S_{\beta,\delta}\overline{S_{\delta,\gamma}}}{S_{o,\delta}}.
\label{effective_Verlinde_formula}
\end{equation}
Indeed, as we see later, these coefficients appear as the coefficient of fusion algebra of symmetry operators or defects, whereas the defect realization requires more careful treatment.
This matrix is less familiar, but it has already appeared in  \cite{Gannon:2001uy,mueger2002structuremodularcategories}, and the section $8.19.$ of  \cite{etingof2015tensor}, and it has been known to be integers. Eq. \eqref{effective_Verlinde_formula} has been studied in some categorical methods, but its realization and interpretation are less clear, because this can be outside of standard NIM-rep.

We demonstrate that there exists the corresponding algebra, $\mathbf{A}^{[o]}=\{ \mathcal{Q}_{\alpha}^{[o]}\}$, which is ring isomorphic to $\mathbf{A}=\{ \mathcal{Q}_{\alpha}\}$. Moreover, the new algebra $\mathbf{A}^{[o]}$ corresponds to the SymTFT of a nonlocal unitary CFT constructed from a local nonunitary CFT through the similarity transformation. Our primary method is linear algebra following the approach in \cite{Petkova:2000ip}. Because of the fundamental theoretical strength of linear alegbra to quantum systems, our method applies to more general pseudo-Hermitian systems in principle. Hence, if there exists a unitary CFT that has the same character as a nonunitary CFT, one can obtain the dual relation between their symmetries. Fortunately, there exists at least two series of  orrespondence dual under the similarity, the correspondence between symplectic fermions and Dirac fermions \cite{Flohr:1996vc,Guruswamy:1996rk} (or related $Sp(2n)_{k}\times Sp(2k)_{n}=SO(4nk)_{1}$ duality \cite{Verstegen:1990at,Aharony:2016jvv}), and those between $M(2,2k+3)$ minimal models and $Osp(1, 2)_{k}$ models (or related $W$-minimal models and $Osp(1,2n)_{k}$ models \cite{Ferrari:2023fez,Creutzig:2024ljv}). We expect the same type of dualities for the models with positive $\{ S_{o,\alpha}\}_{\alpha}$, and we write the corresponding SymTFTs and their modules. To our knowledge, the significance of the condition $\{ S_{o,\alpha}>0\}_{\alpha}$ in nonunitary CFTs has never captured sufficient attention in the community\cite{Gannon:2003de,Fukusumi:2025xrj}, but this condition is fundamental in constructing $\mathbf{A}^{[o]}$.

\section{Characters and Symmetry operators by shuffle}

First, we introduce the shuffle operation \cite{Gannon:2003de} $\eta^{[o]}$, the chiral analog of the similarity transformation \cite{Mostafazadeh:2001jk,Mostafazadeh:2001nr,Mostafazadeh:2002id}, with effective-vacuum $o$ as follows,
\begin{equation}
\eta^{[o]}: \eta^{[o]}(|\alpha, M\rangle)= |\alpha, M\rangle^{[o]}
\end{equation}
with the following shuffled representation of the Virasoro generator and the characters
\begin{equation}
L_{0}^{[o]}-\frac{c_{\text{eff}}}{24}=\eta^{[o]}L_{0}{\eta^{[o]}}^{-1}-\frac{c}{24}
\end{equation}
\begin{equation}
\chi_{\alpha}(\tau)=\sum_{M}\langle\alpha, M|^{[o]} e^{2\pi i \tau (L_{0}^{[o]}-c^{[o]}/24)} |\alpha, M\rangle^{[o]}
\end{equation}
where $h_{\alpha,M}^{[o]}=h_{\alpha,M}-h_{o}$ and $c_{\text{eff}}=c^{[o]}=c-24h_{o}$ are called effective conformal dimensions or effective central charges, respectively. Significance of the above representations have been first noticed in  \cite{Guruswamy:1996rk} and applied to the analysis of gapless fractional quantum Hall states \cite{PhysRevLett.60.956,PhysRevB.75.075317,Davenport:2012fcs,Bernevig_2008} in  \cite{Milovanovic:1996nj,Ino:1998by,Fukusumi_2022,Fukusumi:2024ejk}. For the antichiral parts, we also assume the same shuffle operation and denote it as $\overline{\eta}^{[o]}$.

Whereas we \emph{assumed} the existence of the shuffle operation, a chiral and antichiral analogue of similarity transformations, but the combination $\eta^{[o]} \overline{\eta}^{[o]}$ can rigorously be identified as the similarity tranformation by applying the arguments in  \cite{Mostafazadeh:2001jk,Mostafazadeh:2001nr,Mostafazadeh:2002id}. At the level of linear algebra, the existence of similarity transformations is obvious, but their analytical properties are not studied except for the simplectic fermion  \cite{Guruswamy:1996rk}. Corresponding to the above unitary representation, we also introduce the following projections,
\begin{align}
P_{\alpha,M}^{[o]}=|\alpha, M\rangle^{[o]}\langle\alpha, M|^{[o]}
\end{align}
We have assumed the Hermiticity for the quantum states, $\langle \alpha, M|^{[o]}=\left(|\alpha,M\rangle^{[o]}\right)^{\dagger}$, whereas they are defined by vector and linear dual in nonunitary CFTs. By applying the same procedure for the antichiral parts, we introduce the following idempotent,
\begin{equation}
P_{\alpha}^{[o]}=\sum_{M,\overline{M}}P_{\alpha,M}^{[o]}\overline{P_{\alpha,\overline{M}}^{[o]}}
\end{equation} 
This expression induces the ring isomorphism $\{ P_{\alpha}\}=\{ P_{\alpha}^{[o]}\}$. 

By definition, this shuffle operation implements the following symmetry operators,
\begin{equation}
\mathcal{Q}_{\alpha}^{[o]}=\sum_{\beta} \frac{S_{\alpha,\beta}}{S_{o,\beta}} P_{\beta}^{[o]}
\label{shuffle_SymTFT}
\end{equation}
with $S_{o,\beta}\neq 0$. In many series of nonunitary CFTs, the condition $\{S_{o,\beta} > 0\}_{\beta}$ is satisfied\cite{Gannon:2003de}. Hence, the above symmetry operators are well-defined in many cases, and one can obtain the expression,
\begin{equation}
\mathcal{Q}_{\alpha}^{[o]}\times\mathcal{Q}_{\beta}^{[o]}=\sum_{\gamma}N^{\gamma,[o]}_{\alpha,\beta}\mathcal{Q}_{\gamma}^{[o]}
\end{equation}
by applying the definition of $N^{[o]}$, a generalization of Verlinde formula. Consequently, we have demonstrated the implication of the fusion coefficient defined by $o$, only by assuming the unitarity. We also note that $\mathcal{Q}_{o}^{[o]}=\sum_{o} P_{\beta}^{[o]}$ behaves as the identity in $\mathbf{A}^{[o]}=\{\mathcal{Q}_{\alpha}^{[o]} \}$ and the roles of $I$ and $o$ are \emph{shuffled}. In the next section, we demonstrate the naturalness of the algebra $\mathbf{A}^{[o]}$, by identifying its quantum dimensions and modules (or smeared BCFTs) and by studying their implications to RGs.

It should be stressed that the condition $\{S_{j,\beta}\neq 0\}_{\beta}$ implies the invertibility of $\mathcal{Q}_{j}$ in \emph{linear algebra} (not in category theory). It is well-known that the invertibility of the symmetry plays a significant role in studying gauging operations \cite{Tachikawa:2017gyf,Bhardwaj:2017xup} even in category theory, and our argument provides a generalization respecting the linear algebraic structure of quantum systems. Moreover, one can obtain the following relations,
\begin{equation}
P_{\alpha}^{[o]}=\sum_{\beta}\mathcal{Q}_{\beta}^{[o]} S_{o,\alpha}\overline{S_{\alpha,\beta}}
\end{equation}
Hence, there exists a series of ring isomorphisms,
\begin{equation}
\{ \mathcal{Q}_{\alpha}\}=\{ P_{\alpha}\}=\{ P_{\alpha}^{[o]}\}=\{ \mathcal{Q}_{\alpha}^{[o]}\}
\end{equation}
These relations imply the equivalence between the classifications of RGs (or SymTFTs) in nonunitary CFTs and those in nonlocal unitary CFTs. However, the matrix $N^{[o]}$ can be outside of NIM-rep \cite{Cardy:1986gw,Cardy:1989ir}, and a simple object, such as $\mathcal{Q}_{\alpha}$, cannot be mapped to a simple object in $\mathbf{A}^{[o]}$ in general. For SymTFTs, this aspect does not affect the analysis, but it does the analysis for boundary conditions or defects, where the coefficients before objects need to become nonnegative integers. We also note that our generalization of fusion rules is different from those constructed from gauging operations \cite{Ishikawa:2005ea,Choi:2024tri,Das:2024qdx,Choi:2024wfm}.

\subsection{$M(2,5)$ minimal model and $Osp(1,2)_{1}$ WZW model}
In the $M(2,5)$ minimal model, there exists only two primary fields $I$ and $o$ with the conformal dimensions $h_{I}=0$, $h_{o}=-1/5$, and with the central charge $c=-22/5$ . The modular S matrix is given as follows,
\[
S = \frac{2}{\sqrt{5}}
\begin{pmatrix}
-\sin\left(\frac{2\pi}{5}\right) & \sin\left(\frac{4\pi}{5}\right) \\
\sin\left(\frac{4\pi}{5}\right) & \sin\left(\frac{2\pi}{5}\right)
\end{pmatrix}
\]
The nontrivial part of the fusion ring $\mathbf{A}$ is,
\begin{equation}
\mathcal{Q}_{o}\times \mathcal{Q}_{o}=\mathcal{Q}_{I}+\mathcal{Q}_{o},
\end{equation}
where $\mathcal{Q}_{I}$ is the identity operator. $\{ \mathcal{Q}_{I},\mathcal{Q}_{o}\}$ is called the Fibonacci fusion ring and is famous for the description of a non-abelian anyon \cite{Feiguin:2006ydp,Ardonne:2011wxx}.

By applying the shuffle operation, one can obtain the following fusion ring $\mathbf{A}^{[o]}$,
\begin{equation}
\mathcal{Q}_{I}^{[o]}\times \mathcal{Q}_{I}^{[o]}=\mathcal{Q}_{o}^{[o]}-\mathcal{Q}_{I}^{[o]}
\end{equation}
where $\mathcal{Q}_{o}^{[o]}$ is the identity operator. As can be seen from the above expression, the roles of $I$ and $o$ are shuffled, and their conformal dimensions are $h_{o}^{[o]}=0$, $h_{I}^{[o]}=1/5$. These conformal dimensions and the effective central charge $c_{\text{eff}}=2/5$ match with those of $Osp(1,2)_{1}$ Wess-Zumino-Witten (WZW) model \cite{Ferrari:2023fez,Creutzig:2024ljv}. Hence, the above $\mathbf{A}^{[o]}$ provides a natural expression of SymTFT of the $Osp(1,2)_{1}$ WZW model.

\section{Ring isomorphism and application to RGs}
In this section, we study the consequences of the similarity transformation in SymTFTs combined with massive and massless RG. We denote the ring isomorphism as $\iota: \mathbf{A}\rightarrow \mathbf{A}^{[o]}$ in the ultraviolet(UV) theories. At the algebraic level, the massless RG corresponds to a ring homomorphism (or functor), and this aspect has been studied in the name RG domain wall \cite{Brunner:2007ur,Gaiotto:2012np,Klos:2019axh} or gapped domain wall \cite{Lan:2014uaa,Hung:2015hfa,kong2015boundarybulkrelationtopologicalorders,Kong:2017hcw,Zhao:2023wtg}. On the other hand, the massive RG corresponds to the operation taking module of the unbroken symmetry subring of the original UV theory \cite{Date:1987zz,Saleur:1988zx,Han:2017hdv,Foda:2017vog,Cardy:2017ufe,Numasawa:2017crf,Thorngren:2019iar,Kikuchi:2022gfi,Kikuchi:2023gpj,Fukusumi:2024ejk,Wen:2025xka,Choi:2025ebk,Fukusumi:2025clr,Fukusumi:2025xrj}. 

\subsection{Massless RG flow}
In this subsection, we note the implications of the similarity transformations realized as ring isomorphism. We denote infrared (IR) theories as $\mathbf{A'}$ and $\mathbf{A'}^{[o]}$ and their similarity transformation as $\iota': \mathbf{A}'\rightarrow \mathbf{A'}^{[o]}$.  

First, we assume the massless RG or ring homomorphism $\rho$, $\rho: \mathbf{A}\rightarrow \mathbf{A'}$ \cite{Hung:2015hfa,Wan:2016php,Zhao:2023wtg}. The ring homomorphism is a linear map satisfying the following relations,
\begin{equation}
\rho(\alpha\times\beta)=\rho(\alpha)\times \rho(\beta)
\end{equation}
This relation implies that the fusion operation is deformed smoothly from one theory to another, and these smooth transtion properties have a close connection to the appearance of pseudo-Nambu-Goldstone (NG) modes (or Higgs modes) which should be gapped out gradually under the phase transitions \cite{Kaplan:1983sm,Kaplan:1983fs,Dugan:1984hq,Georgi:1984af} (see also reviews \cite{Panico:2015jxa,Watanabe:2019xul,Brauner:2024juy}). In other words, at the level of linear algebra, the ring homomorphism determines the consequence after gapping out pseudo-NG modes, and the kernel of the map, $\text{Ker}(\rho)$, determines the pseudo-NG modes \cite{Fukusumi:2025clr,Fukusumi:2025xrj,Fukusumi:2025fvb}. We note that the related correspondence between the massless RG and the level-rank duality \cite{Kuniba:1990im,Kuniba:1990zh,Nakanishi:1990hj,Hsin:2016blu,Aharony:2016jvv} or coset construction \cite{Goddard:1984vk,Goddard:1986ee} has been studied in  \cite{LeClair:2001yp,Sfetsos:2017sep,Georgiou:2018gpe}. Related discussions can be seen in some recent references \cite{Cordova:2025eim,Apruzzi:2025hvs}. It is worth noting that the simple object, $\alpha$, in the UV theory can be transformed to a nonsimple object $\sum_{\alpha'}A^{\alpha'}_{\alpha}\alpha'$ in the IR theory, where $A$ is the coefficient. Because we have only assumed a linear algebraic structure of SymTFT, the coefficient can be noninteger \cite{Zhao:2023wtg,Fukusumi:2025clr}, and this is a natural generalization of anyon condensation in  \cite{Bais:2008ni,Kong:2013aya}. In other words, if one does not permit the noninteger coefficient, the anyon condensation contains difficulties in classifying the existing quantum Hall hierarchy observed in experiments \cite{Wang2017TopologicalOF,Mross_2018} realizing the anomaly inflow mechanism\cite{Callan:1984sa,Stone:2012ud,Son:2015xqa,Barkeshli:2015afa}. Moreover, this linear algebraic representation implies that the nonsimple object in UV theory can be simple in IR, and one can see the corresponding phenomena in the RG flow $\{ SU(N)_{1} \}^{k}\rightarrow SU(N)_{k}$, because the abelian SymTFT, $\text{Rep}(\{SU(N)_{1}\}^{k})=\{Z_{N}\}^{k}$, flows to the nonabelian one \cite{Lecheminant:2015iga,Fuji_2017} (see  \cite{Bourgine:2024ycr,Zhang:2024bye,Yutushui_2025,Seo:2026wmq}, for other examples in fractional quantum Hall hierarchies). Related phenomena have been studied in  \cite{Furuya:2015coa,Numasawa:2017crf,Kong:2019cuu,Fukusumi_2022_c,Kikuchi:2022ipr,Fukusumi:2024ejk,Antinucci:2025fjp} with a connection to the Haldane conjecture \cite{Haldane:1982rj,Haldane:1983ru,Wamer:2019oge} or anomaly matching. In other words, the emergence of nonabelian anyons or noninvertible symmetries can be understantood reasonable when assuming their underlying structure of coset constructions or level-rank dualities. Hence, nonsimple objects can be stabilized and simple object can split under the phase transition, and invertibility of objects can change dramatically.

\begin{figure}[!ht]
\centering
\resizebox{0.5\textwidth}{!}{%
\begin{circuitikz}
\tikzstyle{every node}=[font=\LARGE]
\tikzset{>={Stealth[length=8pt,width=8pt]}}

\draw  (-4,10.5) rectangle (5.5,-0.9);
\draw  (9,10.5) rectangle (18.5,-0.9);

\draw[fill=lightgray, fill opacity=0.5] (-3.2,9.0) rectangle (4.7,5.5);   
\draw[fill=lightgray, fill opacity=0.5] (-3.2,3.5) rectangle (4.7,0.0);    

\draw[fill=lightgray, fill opacity=0.5] (-2,8.2) rectangle (3.5,6.3);   
\draw[fill=lightgray, fill opacity=0.5] (-2,2.7) rectangle (3.5,0.8);

\draw[fill=lightgray, fill opacity=0.5] (9.8,9.0) rectangle (17.7,5.5);   
\draw[fill=lightgray, fill opacity=0.5] (9.8,3.5) rectangle (17.7,0.0);    

\draw [<->, line width=1.2pt] (0.75,5.5) -- (0.75,3.8);
\draw [<->, line width=1.2pt] (13.75,5.5) -- (13.75,3.5);

\draw(1.5,4.5) node[right] {\huge $\iota$};
\draw(0.5,4.5) node[left] {\huge Shuffle};
\draw(14.5,4.5) node[right] {\huge $\iota'$};
\draw(13.5,4.5) node[left] {\huge Shuffle};
\draw [->, line width=1.2pt] (3.5,1.75) -- (9.8,1.75);

\draw(7,2) node[above] {\huge $\rho$};
\draw(7,1.5) node[below] {\huge RG};
\draw [->, line width=1.2pt] (3.5,7.5) -- (9.8,7.5);

\draw(7,8) node[above] {\huge $\iota^{'}\rho\iota^{-1}$};
\draw(7,7) node[below] {\huge RG};

\node[draw, rectangle, fill=white, inner sep=6pt, minimum height=1.2em] at (0.75,10.5) {\LARGE UV theory };
\node[draw, rectangle, fill=white, inner sep=6pt, minimum height=1.2em] at (13.75,10.5) {\LARGE IR theory};

\node[draw, rectangle, fill=white] at (0,9.1) {\LARGE$\mathbf{A}^{[o]}$: unitary, non local};
\node[draw, rectangle, fill=white] at (-1.5,8) {\LARGE$\mathbf{A}^{[o]}_{\text{ub}}$};
\node[draw, rectangle, fill=white] at (0.75,6.5) {\LARGE $\mathcal{Q}_{\alpha}^{[o]}\times\mathcal{Q}_{\beta}^{[o]}=\sum_{\gamma}N^{\gamma,[o]}_{\alpha,\beta}\mathcal{Q}_{\gamma}^{[o]}$};
\node[draw, rectangle, fill=white] at (0,3.5) {\LARGE $\mathbf{A}$:non unitary, local};
\node[draw, rectangle, fill=white] at (-1.5,2.5) {\LARGE $\mathbf{A}_{\text{ub}}$};

\node[draw, rectangle, fill=white] at (0.75,1) {\LARGE $\mathcal{Q}_{\alpha}\times\mathcal{Q}_{\beta}=\sum_{\gamma}N^{\gamma}_{\alpha,\beta}\mathcal{Q}_{\gamma}$};

\node[draw, rectangle, fill=white] at (11.5,9) {\LARGE${\mathbf{A}'}^{[o]}$};
\node[draw, rectangle, fill=white] at (13.75,6.5) {\LARGE ${\mathcal{Q}^{[o]}}'_{\alpha}\times{\mathcal{Q}^{[o]}}'_{\beta}=\sum_{\gamma}{N'}^{\gamma,[o]}_{\alpha,\beta}{\mathcal{Q}'}^{[o]}_{\gamma}$};
\node[draw, rectangle, fill=white] at (11.5,3.5) {\LARGE$\mathbf{A}'$};

\node[draw, rectangle, fill=white] at (13.75,1) {\LARGE $\mathcal{Q}'_{\alpha}\times\mathcal{Q}'_{\beta}=\sum_{\gamma}{N'}^{\gamma}_{\alpha,\beta}\mathcal{Q}'_{\gamma}$};
\end{circuitikz}
}%
\caption{The schematic correlation chart of the massless RG flow from UV theory to IR theory. We use "$'$" symbol for IR rings. Notice that the RG flow is generally complicated in the sense that the simple objects in UV can be transformed to some complicated objects in IR. }
\label{fig:massless_flow}
\end{figure}
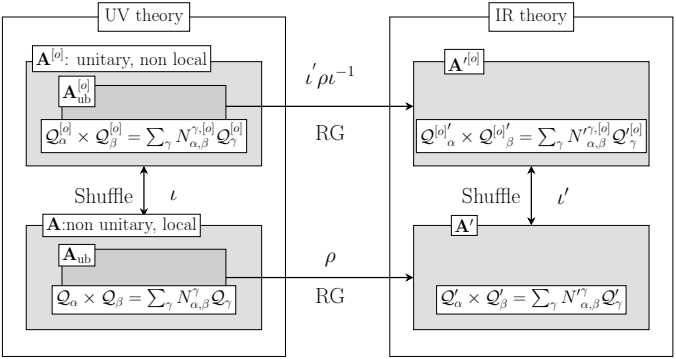

In this setting, based on the above linear algebraic understanding, one can implement the ring homomorphism of the corresponding nonlocal unitary model as 
\begin{equation}
\iota' \rho\iota^{-1}: \mathbf{A}^{[o]}\rightarrow \mathbf{A'}^{[o']} 
\end{equation}
The unbroken symmetry $\mathbf{A}_{\text{ub}}$, such as $\mathbf{A}_{\text{ub}}\cap \text{Ker}\rho =\{ 0\}$ in the UV theory is transformed to $\iota (\mathbf{A}_{\text{ub}})$ and one can apply the analysis in  \cite{Fukusumi:2025clr,Fukusumi:2025xrj}. See FIG. \ref{fig:massless_flow} for intuition. The theoretical definition of the unbroken symmetry is straightforward \cite{Fukusumi:2025clr}, but its detection in particular massless flows is still in development. We note several recent works  \cite{Nakayama:2024msv,Kikuchi:2024cjd,Gaberdiel:2026sfg,Ambrosino:2026umb}. However, we also point out that, whereas the detection of the unbroken symmetry in massless flows is important, this information cannot fix the corresponding homomorphism uniquely \cite{Fukusumi:2025clr} without some other theoretical assumptions. For readers interested in the concrete descreptions of related phenomena, we note  literature studying related domain wall or junction problems \cite{Quella:2006de,Kimura:2014hva,Brunner:2015vva,Kimura:2015nka,Stanishkov:2016pvi,Stanishkov:2016rgv,Poghosyan:2022ecv,Poghosyan:2022mfw,Cogburn:2023xzw}.

For further studies, we introduce the following generalization of quantum dimensions (which can be called shuffled quantum dimensions),
\begin{equation}
q_{\alpha,(a)}^{[o]}=\frac{S_{\alpha,a}}{S_{o,a}}.
\end{equation}
These quantum dimensions satisfy the relation $q_{\alpha,(a)}^{[o]}\times q_{\beta,(a)}^{[o]}=\sum_{\gamma}N^{\gamma,[o]}_{\alpha,\beta}q_{\gamma,(a)}^{[o]}$. Hence, one can implement the ring homomorphism $d_{(a)}^{[o]}:\mathbf{A}\rightarrow \mathbb{C}$ as $d_{(a)}^{[o]}(\alpha)=S_{\alpha,a}/S_{o,a}$.

Interestingly, the simple objects in $\mathbf{A}_{\text{ub}}$ are not mapped to simple objects in $\iota (\mathbf{A}_{\text{ub}})$ in general. By combining this observation with the homomorphism $\rho$, one can notice that the nonsimple (or nonlocal) objects in $\mathbf{A}$ can be simple objects in the IR theory $\mathbf{A'}_{\text{ub}}^{[o]}$. At the level of linear algebra, this phenomenon is understandable, but its phenomenologies or categorical descriptions are less clear. 

\subsection{Smeared BCFT and massive RG flow} 
Corresponding to the representation of the symmetry operators, one can implement the following as smeared Cardy states $\{ |\alpha\rangle^{[o]} \}$ and their \emph{complex conjugations} $\{ \langle\alpha|^{[o]}\}$,
\begin{equation}
|\alpha\rangle^{[o]} =\sum_{\beta}\frac{S_{\alpha,\beta}}{\sqrt{S_{o,\beta}}}|\beta \rangle \rangle^{[o]}, \langle\alpha|^{[o]} =\sum_{\beta}\frac{\overline{S}_{\alpha,\beta}}{\sqrt{S_{o,\beta}}}\langle \langle \beta |^{[o]}
\label{shuffle_BCFT}
\end{equation}
where the Ishibashi states $\{ |\alpha\rangle\rangle\}$ satisfy the relation $\langle\langle \alpha|^{[o]} e^{\pi i\tau H} |\beta \rangle \rangle^{[o]}=\chi_{\alpha}(\tau)\delta_{\alpha,\beta}$. We stress the fundamental significance of the condtion $\{ S_{o,\alpha}>0\}_{\alpha}$ in constructing the bra and ket by the complex conjugation, in comparison with nonunitary CFTs.

These boundary states can be interpreted as module of $\mathbf{A}^{[o]}=\{ \mathcal{Q}_{\alpha}^{[o]} \}$, satisfying the following relation \cite{Graham:2003nc},
\begin{equation}
\mathcal{Q}_{\alpha}^{[o]}|\beta\rangle^{[o]}=\sum_{\gamma}N^{\gamma,[o]}_{\alpha,\beta}|\gamma\rangle^{[o]} \ \left(=|\alpha\times \beta \rangle^{[o]}\right).
\end{equation}
where we call the right-hand side as the Graham-Watts (GW) states. Hence, by assuming the massive RG is realized as reduction from $\mathbf{A}^{[o]}$ to $\mathbf{A}_{\text{ub}}^{[o]}$ \cite{Cardy:2017ufe,Thorngren:2019iar,Kikuchi:2022gfi,Kikuchi:2023gpj,Fukusumi:2024ejk,Fukusumi:2025clr,Fukusumi:2025xrj}, one can classify the gapped phase as $\mathbf{A}_{\text{ub}}^{[o]}$ invariant subspace of the Hilbert space spaned by $\{ |\alpha\rangle^{[o]} \}$. 
In this section, the unitarity combined with recent developments in BCFT \cite{Graham:2003nc,Cardy:2017ufe} has played a fundamental role, and the existing arguments in the \emph{smeared} BCFT is applicable to our cases. However, the Cardy's condition which determines the \emph{boundary critical phenomena} is broken, because the amplitude can become outside of NIM-rep \cite{Cardy:1989ir},
\begin{equation}
\begin{split}
Z_{\alpha,\beta}^{[o]}(\tau)&=\langle \alpha |^{[o]}e^{\pi i\tau (L_{0}^{[o]}+\overline{L_{0}}^{[o]}-c_{\text{eff}}/12)}|\beta\rangle^{[o]} \\
&=\sum_{\gamma}N^{\gamma,[o]}_{\alpha,\beta}\chi_{\gamma}(-1/\tau)
\end{split}
\end{equation}
It is reasonable because smeared BCFTs correspond to the quantum states of gapped phases, and the amplitudes correspond to the inner products of quantum states, which take values in $\mathbb{C}$  \cite{Fukusumi:2025clr,Fukusumi:2025ljx,Fukusumi:2025xrj}.
Note that the state $\ket{\alpha}^{[o]}$ is not a boundary state for $H=L_0+\overline{L}_0-\frac{c}{24}$ in the sense that they are not satisfying the Cardy condition for $L_n$ and $\overline{L}_n$.  Corresponding to this fact the amplitude includes boundary entropy $\bra{I}\ket{\alpha}^{[o]}=\log{g_{\alpha}^{[I;o]}}$ in low temperature limit $\tau = i \beta, \beta \to 0$, whereas it becomes positive by defining $\bra{o}\ket{\alpha}^{[o]}=\log{g_{\alpha}^{[o;o]}}$. 

In a local theory, the GW states can be identified as the local boundary conditions generated by the Cardy boundary condition with some additional local degrees of freedom \cite{Fukusumi:2020irh}. This is a consequence of the locality of boundary conditions and defects which are generated from the fusion of simple objects $a\times b \times c ...$. The GW states provide a well-organized set of boundary conditions and the corresponding amplitudes in a local theory. But this is not the case for the nonlocal theory, and we discuss the problem in Sec. \ref{Difficulty_BCFT}. In other words, the insertion of symmetry operators and that of defects are intrinsically different in nonlocal systems, because the latter is constrained by the positive amplitude condition. The distinction between the symmetry operators and defects, called emanant symmetry, has been studied in  \cite{Grimm:1990ch,Grimm:2001dr,Belletete:2018eua,Belletete:2020gst,Seiberg:2023cdc}, but our case is intrinsically different because the large volume limit does not resolve the discrepancies.

\section{Difficulty in boundary or defect phenomena}
\label{Difficulty_BCFT}
As we have observed in the previous section, the ring $\mathbf{A}^{[o]}$ outside of the NIM-rep \cite{Cardy:1986gw,Cardy:1989ir} cannot describe the defect or boundary phenomena straightforwardly. However, there still exist similarity transformations for the non-Hermitian system with such objects. For simplicity, we concentrate our attention on the annulus nonunitary CFT determined by $Z_{\alpha, \beta}$. In this setting, by the open-closed duality \cite{Cardy:1989ir}, one can map the Hilbert space to that of the chiral CFT, $\oplus_{\gamma} N^{\gamma}_{\alpha,\beta}\gamma$. Hence, we propose a conjecture for the shuffle $\eta^{[o]}$ and the similarity transformation:  
\begin{equation}
\text{$\eta^{[o]}$ implements the similarity transformation to $\oplus_{\gamma} N^{\gamma}_{\alpha,\beta}\gamma$. }
\end{equation}

Under this assumption, one can apply the existing formalisms in the chiral CFTs to the unexplored boundary physics of the nonlocal models. More technically, the doubling trick \cite{PhysRevLett.54.1091,Bloete:1986qm} will play a fundamental role in studying the correlation functions and related quantities. We also point out that the above similarity transformation induced by some invertible operator can be regarded as a generalization of the large gauge transformation induced by unitary operators, known as Lieb-Schultz-Mattis operators \cite{Lieb:1961fr,Schultz:1964fv,Oshikawa_2000,Aligia_2005}\footnote{We thank Masaki Oshikawa for the corresponding discussion.}. This kind of interpretation has appeared in  \cite{Fernandez:2015aqe}.

By choosing the left and right boundary conditions carefully for $Z_{\alpha,\beta}^{[o]}$, one can also avoid the negative amplitude. This nontrivial (or nonlocal) relation between the nonnegative amplitude condition and the boundary conditions can be interpreted as a signal of the nonlocality. For example, in the $M(2,5)$ model, one can obtain the same characters from the pairs $(Z_{o,o}^{[o]},Z_{I,o})$, $(Z_{o,I}^{[o]},Z_{I,I})$, but it is impossible to reproduce $Z_{o,o}$ from $Z^{[o]}_{\alpha,\beta}$, by taking the label $\alpha, \beta$ as that of the GW states. By introducting the unusual non-GW states $|o\rangle^{[o]}+|I\rangle^{[o]}$, $Z_{o,I+o}^{[o]}$ produces $Z_{o,o}$, but this breaks the correspondence between left and right boundaries in $Z_{o,o}$. In other words, by constructing a nonlocal Hermitian system from a local pseudo-Hermitian system, there can exist unconventional boundary conditions outside of the GW states in nonlocal unitary BCFT. Inversely, there exists some cases where a nonlocal system with non-GW states can be mapped to a pseudo-Hermitian system with GW states.

The phenomenon seems to be related to the boundary sensitity \cite{Edvardsson:2019juv,Koch:2020yfq,Edvardsson:2022bfu} or quantum skin effect \cite{Yao:2018fsg} in non-Hermitian systems, and it is worth further study. We also note some works studying similar problems arising from the boundary conditions of pseudo-Hermitian systems \cite{Morin-Duchesne:2015afa,Kawabata:2018edo,Chou:2025awd}, and a pioneering work pointing out its relation to quantum groups \cite{Pasquier:1989kd}.

\section{Conclusion and discussion}
In this paper, we have studied the implication of the Galois shuffle operations in CFTs \cite{Gannon:2003de} and their implications as similarity transformations between pseudo-Hermitian systems and the corresponding nonlocal Hermitian systems. Essentially, this is a ring isomorphism between two theories, and the RGs of one side are compatible with the other side. However, when studying the implications for defects or boundaries, more careful treatment is necessary, and this difficulty is a natural consequence of the nonlocality. We remind that our discussion is based on linear algebra fundamental in studying quantum mechanical systems, and this aspect implies the necessity studying $\mathbb{C}$-linear category compatible with linear algebra.

As a future application, our method may have a useful tool to analyse dS/CFT. In  \cite{Strominger:2001pn}, homolographic CFTs dual to dS spacetime are given by non-unitary CFTs with certain complex central charges. Possibly, similar to the similarity transformation, the CFTs are obtained by analytic continuation from one dual to AdS. Similarly to $\ket{\alpha}^{[o]}$, we observed some boundary states that have complex boundary entropy that are dual to dS branes in the bulk  \cite{Akal:2021dqt,Hao:2024nhd}. The understanding of these exotic behaviours related to cosmological settings an interesting future direction.

\section{Acknowledgement}
The authors thank Naomichi Hatano, Hosho Katsura, and Po-Yao Chang for sharing their knowledge on the similarity transformations. YF thanks Yuma Furuta and Shinichiro Yahagi for related collaborations. YF also thanks Ingo Runkel and Sylvain Ribault for the discussions on fusion rules and Jurgen Fuch and Yuji Tachikawa for the helpful discussions on the fusion ring. We thank Masaki Oshikawa for the helpful discussions on the large gauge transformation. YF thanks Shumpei Iino for the discussions on non-GW states. We thank Chen-Te Ma for the helpful comments on the literature on the realization of chiral fermions in non-Hermitian systems. TK is supported by Grant-in-Aid for JSPS Fellows No. 23KJ1315. YF tthe hanks support from NCTS.

\appendix

\section{$M(2,7)$ minimal models and $Osp(1,2)_{2}$ WZW model}

In this subsection, we summarize the fusion coefficients of the $M(2,7)$ minimal model $N$ and those of the $Osp(1,2)_{2}$ WZW model, $N^{[o]}$. The $M(2,7)$ minimal model has three primary fields, $\{I, \phi, o\}$ with the conformal dimensions $\{h_{I}=0, h_{\phi}=-2/7,h_{o}=-3/7\}$. The conformal dimension and the effective conformal dimension are $c=-68/7$ and $c_{\text{eff}}=4/7$. Hence, by considering the effective central charge $c_{\text{eff}}$ and effective conformal dimensions $h^{[o]}_{\alpha}$, one can check its correspondence to $Osp(1,2)_{2}$ WZW model. The modular $S$-matrix for this model is,
\begin{equation}
S = \frac{2}{\sqrt{7}}
\begin{pmatrix}
 \sin \frac{2\pi}{7} & -\sin \frac{4\pi}{7} & \sin \frac{6\pi}{7} \\
 -\sin \frac{4\pi}{7} & \sin \frac{8\pi}{7} & -\sin \frac{12\pi}{7} \\
 \sin \frac{6\pi}{7} & -\sin \frac{12\pi}{7} & \sin \frac{18\pi}{7}
\end{pmatrix}.
\end{equation}
One can construct $\{\mathcal{Q}_{\alpha}\}$ and $\{ \mathcal{Q}_{\alpha}^{[o]} \}$ from these modular data. We summarize the corresponding fusion coefficients in the following table.

\begin{center}
\begin{tabular}{cccccc}
\hline
$\alpha$ & $\beta$ & $\gamma$ & ${N^{\gamma,[I]}}_{\alpha\beta}$ & ${N^{\gamma,[o]}}_{\alpha\beta}$ \\
\hline
$I$ & $I$ & $I$ & $1$ & $0$ \\
$I$ & $I$ & $\phi$ & $0$ & $-1$ \\
$I$ & $I$ & $o$ & $0$ & $1$ \\
$I$ & $\phi$ & $I$ & $0$ & $-1$ \\
$I$ & $\phi$ & $\phi$ & $1$ & $1$ \\
$I$ & $\phi$ & $o$ & $0$ & $0$ \\
$I$ & $o$ & $I$ & $0$ & $1$ \\
$I$ & $o$ & $\phi$ & $0$ & $0$ \\
$I$ & $o$ & $o$ & $1$ & $0$ \\
$\phi$ & $I$ & $I$ & $0$ & $-1$ \\
$\phi$ & $I$ & $\phi$ & $1$ & $1$ \\
$\phi$ & $I$ & $o$ & $0$ & $0$ \\
$\phi$ & $\phi$ & $I$ & $1$ & $1$ \\
$\phi$ & $\phi$ & $\phi$ & $0$ & $-1$ \\
$\phi$ & $\phi$ & $o$ & $1$ & $1$ \\
$\phi$ & $o$ & $I$ & $0$ & $0$ \\
$\phi$ & $o$ & $\phi$ & $1$ & $1$ \\
$\phi$ & $o$ & $o$ & $1$ & $0$ \\
$o$ & $I$ & $I$ & $0$ & $1$ \\
$o$ & $I$ & $\phi$ & $0$ & $0$ \\
$o$ & $I$ & $o$ & $1$ & $0$ \\
$o$ & $\phi$ & $0$ & $0$ & $0$ \\
$o$ & $\phi$ & $\phi$ & $1$ & $1$ \\
$o$ & $\phi$ & $o$ & $1$ & $0$ \\
$o$ & $o$ & $I$ & $1$ & $0$ \\
$o$ & $o$ & $\phi$ & $1$ & $0$ \\
$o$ & $o$ & $o$ & $1$ & $1$ \\
\hline
\end{tabular}
\end{center}

More generally, one can observe the matching of the effective central charge $c_{\text{eff}}=2k/(2k+3)$ of  $M(2,2k+3)$ minimal model and the central charge $c$ of the $Osp(1,2)_{k}$ WZW model. Phenomenologically, the $M(2,2k+3)$ minimal model corresponds to the bosonic-local-nonunitary CFT, and the $Osp(1,2)_{k}$ WZW model corresponds to the fermionic-nonlocal-unitary CFT. By interpreting the $M(2,2k+3)$ minimal model as (multicritical) Lee-Yang model \cite{Lencses:2022ira} and the $Osp(1,2)_{k}$ WZW model as a supersymmetric model, this is analogous to dimensional reduction (or its breaking) \cite{Parisi:1979ka}. The representation theories and fusion constants of the $Osp(1,2)_{k}$ WZW model are discussed in  \cite{Ennes:1997vt,Fan:1993ps}.
We also note that the correspondence between the fermionic and bosonic representations of the characters is known as the Andrews-Gordon type identity \cite{0265a037-3ee6-3e2f-854a-07dfe87749bf,095af682-15e6-359d-ac8f-df779afaf867}, a particular type of the Rogers-Ramanujan identities (see also  \cite{Nahm:1992sx,Warnaar_2023,Campbell_2024} and literature therein). More recently, related phenomena in Chern-Simons theories have been studied in  \cite{Hsin:2016blu,Aharony:2016jvv}. Related phenomenologies have been summarized in  \cite{Fukusumi:2024nho} by the first author, and we only note a few earlier works \cite{Bershadsky:1989tc,Saleur:1989gj,1989LMaPh..18..143K,Saleur:2001cw} here.

\section{Generalization and conjecture on similarity transformation}

In this section, we comment on a straightforward generalization of our discussion by replacing the symbol $[o]$ to the general symbol $[j]$, satisfying the relation $S_{j,\alpha}\neq 0$ for all $\alpha$. This condition itself is nothing but the invertibility condition of the operator $\mathcal{Q}_{j}$ or anyon $j$ in the fusion ring. Hence, we conjecture 
\begin{equation}
\text{Invertible anyon $j$} \sim \text{similarity transformation $\eta^{[j]}$}
\end{equation}
However, we note that the $\mathcal{Q}_{j}$ itself does not implement the similarity transformation because it commute with Hamiltonian. Because one can apply the method in the main text only by replacing the symbols (and complex conjugate to linear dual), we only note the fusion coefficients:  
\begin{equation}
N^{\gamma, [j]}_{\alpha,\beta}=\sum_{\delta}\frac{S_{\alpha,\delta}S_{\beta,\delta}\overline{S_{\delta,\gamma}}}{S_{j,\delta}}.
\label{invertible_Verlinde_formula}
\end{equation}

In Hermitian systems, it is known that the Lieb-Schultz-Mattis operators $U_{\alpha}$, which can be realized as a unitary operation changes or shuffle the energy eigenstates. Moreover, they correspond tothe  primary field $\alpha$ and generate many-body excitation from the ground state $U_{\alpha}|0\rangle= \alpha(0)|0\rangle=|\alpha\rangle$ where we have assumed the radial quantization in CFTs and $|0\rangle$ is the vacuum. This unitary operation is called a large gauge transformation, and the analogy between the similarity transformation and the gauge transformation has already been pointed out in \cite{Fernandez:2015aqe}. Hence, we conjecture that the similarity transformation can be implemented as invertible operator $U_{j}$ corresponding to the invertible anyon $j$ or the many-body excitation via the CFT expression $U_{j}|0\rangle= j(0)|0\rangle =|j\rangle$ where $j(0)$ corresponds to a generalized version of LSM operator $U_{j}$ generating the many-body excitations \cite{Lieb:1961fr,Schultz:1964fv,Oshikawa_2000,Aligia_2005}.

\bibliographystyle{ytphys}
\bibliography{nonhermitian}

\end{document}